# Dynamic scanning probe microscopy of adsorbed molecules on graphite

N. Berdunov, A.J. Pollard and P.H. Beton

*School of Physics & Astronomy, University of Nottingham, Nottingham NG7 2RD, UK*

We have used a combined dynamic scanning tunneling and atomic force microscope to study the organisation of weakly bound adsorbed molecules on a graphite substrate. Specifically we have acquired images of islands of the perylene derivative molecules. These weakly bound molecules may be imaged in dynamic STM, in which the probe is oscillated above the surface. We show that molecular resolution may be readily attained and that a similar mode of imaging may be realised using conventional STM arrangement. We also show, using tunnelling spectroscopy, the presence of an energy gap for the adsorbed molecules confirming a weak molecule-substrate interaction.

Shortly after the invention of scanning probe microscopy the imaging of adsorbed molecules emerged as a major theme of research which has since grown to encompass manipulation and self assembly phenomena and is also highly relevant to organic electronics [1-4]. While the initial focus was on the use of scanning tunneling microscopy (STM) to acquire images with molecular resolution, there has recently been great progress in applying atomic force microscopy (AFM) to the imaging of adsorbed molecules [5]. Images have now been acquired by a number of groups using either cantilevers or quartz tuning forks as force sensors [6,7]. Within this body of work the approach of Giessibl and co-workers is particularly attractive since it is well suited to the simultaneous measurement of tunnel currents and force gradients [8-10]. In addition this approach has led to different imaging modes such as dynamic STM [10] in which the time averaged current flowing between a sample of interest and a metallic tip attached to an oscillating quartz tuning fork is used as the feedback signal.

In this paper we show that it is possible to resolve single molecules using dynamic STM and also demonstrate additional attractive features such as improved stability, as compared with conventional STM. We also show that these advantages may be realised using a conventional STM tip. For the molecule studied, perylene tetra-carboxylic di-imide (PTCDI), adsorbed on graphite we also observe a reproducible contrast inversion in the dynamic mode of STM.

The instrument we have constructed and used for our experiments is a scanning probe microscope based on tuning fork (TF) force sensor [7] which is operated under ultra-high vacuum (UHV) conditions at room temperature. The instrument can operate in three different regimes in conjunction with the TF sensor (spring constant $k = 1800$ N/m):

dynamic AFM in constant frequency mode (feedback signal: frequency shift);

conventional (dc) STM in constant current mode (feedback signal: tip-sample current);

dynamic (ac) STM in constant average current mode (feedback signal: time averaged tip-sample current). For dynamic STM/AFM measurements the TF is mechanically excited using a small segment of the same piezo tube used for x,y,z movement of the tuning fork. For a metallic tip we use an electrochemically-etched 25 µm diameter PtIr wire glued at the end of the TF, and cleaned in UHV by Ar ion sputtering. After attachment of the tip the resonant frequency (measured in UHV) of the TF, $f_0$, is in the range 25-32kHz with a Q-factor of 2000-5000. The tip is electrically isolated from the TF electrodes and connected to the tunnel current pre-amplifier by a separate electrical connection. The bias voltage is applied to the sample. Resolution of tunnel currents down to 10 pA and tip-sample force interaction less than $10^{-10}$ N (calculated using the approach of Sader and Jarvis [11]) have been attained using TF oscillation amplitudes of 0.1-0.5 nm.

Our experimental arrangement also permits the attachment of a conventional STM tip (we use cut PtIr wire), rather than the TF sensor, to the piezoelectric scanner. Dynamic STM is also possible in this configuration and can be realised by applying an sinusoidal signal to the piezoelectric scanner in order to induce an oscillatory motion of the tip (frequency 29kHz, amplitude 0.2-0.3 nm) perpendicular to the surface (defined as the *z* direction). In our arrangement this voltage is applied to the electrode segment which is normally used to induce mechanical excitation of the TF sensor. However a similar arrangement could be used in conjunction with any STM by adding a small oscillatory signal to the voltage controlling the *z* position of the tip. In all cases GXSM software [12] is used for control/data acquisition and WSXM for image processing [13].

A highly oriented pyrolytic graphite (HOPG) sample was cleaved in air and transferred immediately into a UHV chamber. PTCDI was sublimed onto the HOPG substrate (held at room temperature) at an approximate deposition rate of 0.2 ML/min. Deposition of PTCDI results in the formation of highly facetted rod-like islands with typical widths of ~6nm. These islands grow across step edges and form branched structures as imaged in the AFM mode of operation (Fig.1a). Second layer formation starts even for submonolayer coverage indicating a Volmer-Weber growth mode which might be expected for a passive substrate such as graphite, and has also been observed in earlier reports of PTCDI growth on $MoS_2$ [14]. Interestingly, while there have been many studies of adsorbed molecules on graphite at a liquid/solid interface [15], there have been relatively few STM studies performed under vacuum conditions and many of the published images were acquired at low temperature [16-18]. One reason for this is the relatively weak adsorbate-graphite interaction which results both in rapid diffusion of adsorbates and also gives rise to the potential for damage of any overlayers through interaction with the STM tip.

STM imaging (dc mode) of the PTCDI islands often results in island disruption, presumably due to tip-molecule interactions (see above). Such tip-induced surface modification is demonstrated in Fig.1b where the double-layer island is significantly disrupted following several consecutive scans using a tip mounted on a TF sensor in dc-STM mode. However, images of PTCDI islands can be readily acquired using dynamic STM. We re-iterate that in this mode a time averaged tunnel current between sample surface and oscillating tip is used to control probe height. Fig.1e shows a topographic image acquired in dynamic STM mode using a TF mounted tip.

Unexpectedly, the PTCDI islands are imaged as a depression of approximately 0.1nm in contrast to a 0.14nm protrusion in AFM and conventional STM measurements. Simultaneously with topographic dynamic STM imaging we acquire the frequency shift of the TF oscillation, $df$, and the a.c. component, with frequency $f_0$, of the tunnel current. Negative $df$ values correspond to a positive force gradient. The $df$ image in Fig.1e inset shows an increase of tip-surface attractive interaction when the tip is positioned on top of the PTCDI island, consistent with the topography image Fig.1e. The a.c. component of the tunnel current (Fig.1f) does not show a significant contrast variation between bare graphite and PTCDI.

A high resolution dynamic STM image in Fig.1g reveals a close-packed arrangement of PTCDI with a unit cell of 1.5 nm x 1.8 nm. The elongated shape of the islands can be explained by anisotropy of intermolecular interactions, which arises from the interaction of imide and carbonyl groups on neighbouring molecules as has previously been observed for PTCDI and related molecules [19,20].

To determine the origin of the negative contrast in Figure 1, we have acquired a set of images of PTCDI islands using dynamic STM with a conventional (non TF) tip for several different applied bias voltages. These images show that for large absolute bias voltages (1.5 and –1.7V) a positive contrast is observed (Figs 2a and 2d respectively) whereas for smaller absolute bias (0.8V and -1.0V) negative contrast is observed similar to that observed in Fig. 1. For the lower absolute voltages it is not possible to acquire images in dc-STM. Also shown in Figure 2 is an I(V) curve acquired with the tip positioned (statically) over either the graphite or a PTCDI island. A tunneling gap is clearly present over the PTCDI molecules and this low tunnel current region accounts for the negative contrast that we observe. The large tunneling gap observed when the tip is placed above a molecule indicates that, as expected, the electronic coupling between molecule and surface is rather weak.

To investigate further the contrast in dynamic STM we compare in Figure 3 the variation of frequency shift (*df*) and average current ($\langle I_t \rangle$) with *z,* the tip position, acquired above either the bare graphite or a PTCDI island. There is noticeable broadening of the dependence of *df* on *z* (tip position) curve over a PTCDI island. In addition, the $\langle I_t \rangle$ vs. *z* curve has a much less rapid rise as compared with the equivalent curve acquired on the bare HOPG surface.

To model the average current vs. distance dependence we use the common approximation for tunnel current dependence on work function for conventional STM [21], which leads for dynamic STM, to the following expression for average tunnel current,

$$\langle I_t \rangle = I_0 \int_0^{1/f_0} \exp\left[-2a\varphi^{1/2}(z + A_{osc}\sin(2\pi f_0 t))\right] dt = I_0 \exp\left[-2a\varphi^{1/2}z\right] \cdot J_0\left(2aA_{osc}\varphi^{1/2}\right) \quad (2)$$

where $J_0$ is the modified Bessel function of the first kind, $A_{osc}$ is the amplitude of the tip oscillation, $\varphi$ is the work function and $a = (2m)^{1/2}/\hbar$. The exponential dependence of $\langle I_t \rangle$ on *z* is consistent with our experimental measurements (Fig. 3).

Overall our results show that in regions where contrast inversion occurs, which might be expected for weakly bound molecules, image acquisition using conventional dc-STM is difficult, if not impossible, but images can be acquired, with molecular resolution using dynamic STM. We argue that the reason for the enhanced imaging stability is the oscillatory motion of the STM tip which means that when in close proximity to the molecules the tip is moving perpendicular to the surface. A consequence of this is that as the tip passes over the molecular islands (including the edges) the lateral forces which are generated between tip and molecule are expected to be much lower than those which occur for conventional imaging (and are well known to lead to molecular displacement across surfaces - see for example reference [22]). The relative stability of dynamic STM may be considered analagous to the enhanced stability of tapping mode, as compared with contact-mode AFM. The curves shown in Figure 3 support the presence of significant tip-molecule forces through the broadening of the minimum of the force-gradient versus height curve.

In conclusion, we have demonstrated that dynamic STM may be used to acquire structural and electronic measurements for molecular assemblies weakly bound to a substrate. For the case of PTCDI nano-sized islands on a graphite surface the stability and resolution which may be attained in dynamic STM is significantly better than under dc-STM. Furthermore, we demonstrate that dynamic STM may be realised in a conventional (non-TF) STM and that his imaging technique therefore has the potential to be applied to a wide range of materials where stable image acquisition is not possible using dc-STM.

Acknowledgements
This work was supported by the UK Engineering and Physical Sciences Research Council under grant GR/C534158/1. We are very grateful to R.J. Chettle and D.J. Laird for technical support in instrument development and thank N.R. Champness, P. Moriarty and C.J. Mellor for helpful discussions.

**Figure Captions**

**Figure 1.** Images of PTCDI islands on HOPG surface acquired with TF sensor: (a) AFM topography ($df = -3$ Hz; $A_{osc} = 0.5$ nm); (b) –(d) dc STM images showing islands grown on graphite; regions of the needle like growth shown in the top region of (b) are disrupted by successive scanning – see (c) and (d) which show removal of molecules from the island edge ($I_t = 80$ pA, $V_{bias} = 2.0$V); (e) dynamic STM image showing negative height contrast of ~0.1 nm per layer ($<I_t> = 30$ pA, $V_{bias} = -0.9$V, $A_{osc} = 0.5$ nm); (inset) simulataneously acquired frequency shift (brighter features correspond to higher attractive forces); (f) ac-component of the tunnel current recorded simultaneously with (e); (g) shows close-packed arrangement within the island (unit cell of 1.5 nm x 1.8 nm is marked); (inset) 2D Fourier image;

**Figure 2.** (a) – (d) Dynamic STM images of PTCDI islands acquired using a conventional STM tip ($<I_t> = 30$ pA, $A_{osc} = 0.2$ nm). Images acquired using sample biases (a) 1.8V, (b) 0.8V, (c) -1.0V, (d) -1.7V to illustrate the bias dependence of the negative contrast of the PTCDI islands. (e) Tunneling spectroscopy acquired in dynamic STM mode above the graphite surface and over a PTCDI island illustrating the band gap of the adsorbed molecules.

**Figure 3.** Frequency shift (black) and tunnel current (red) vs. distance measured using a TF sensor on bare HOPG (a) and above a PTCDI island (b). Both approach and withdraw curves are shown. Horizontal axes are given in absolute Z position values.

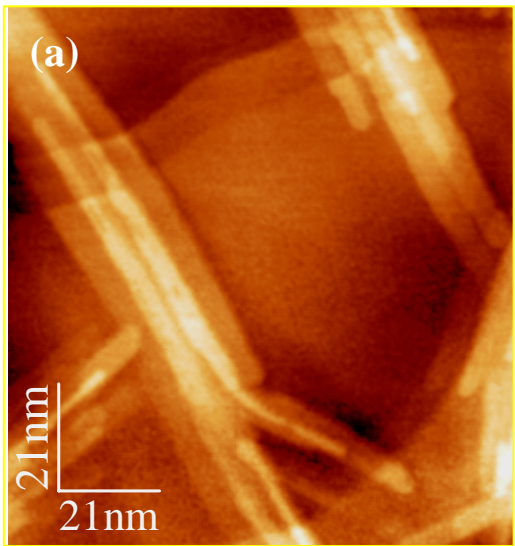
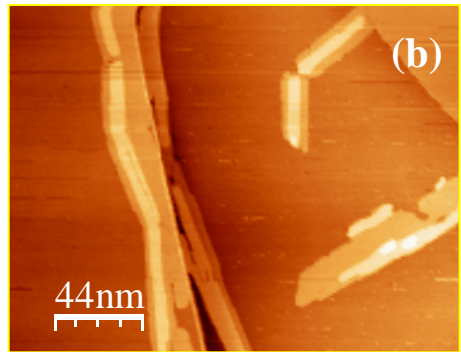
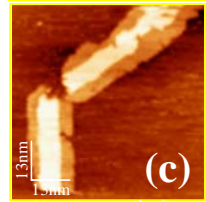
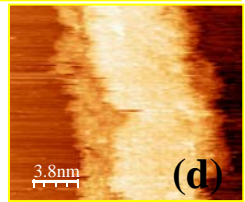
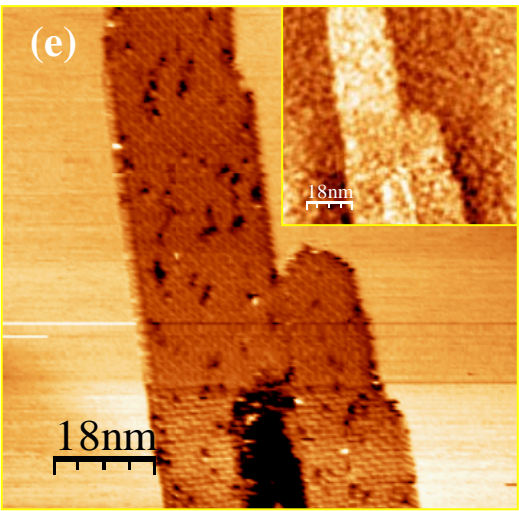
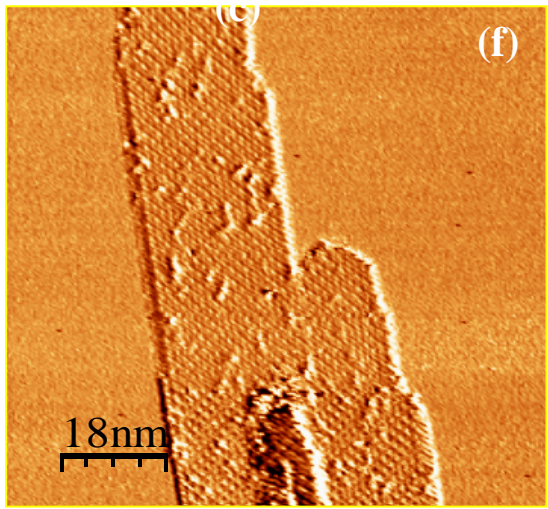
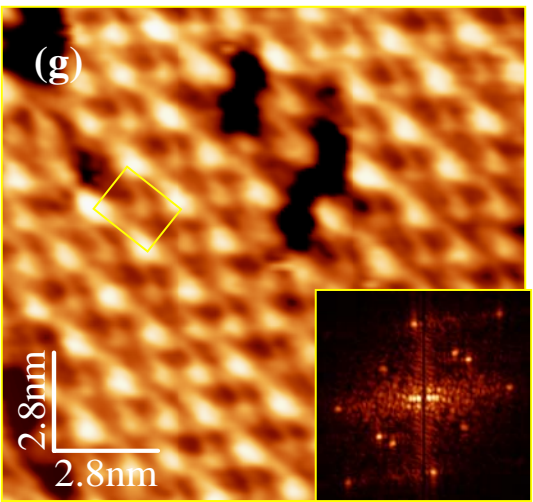

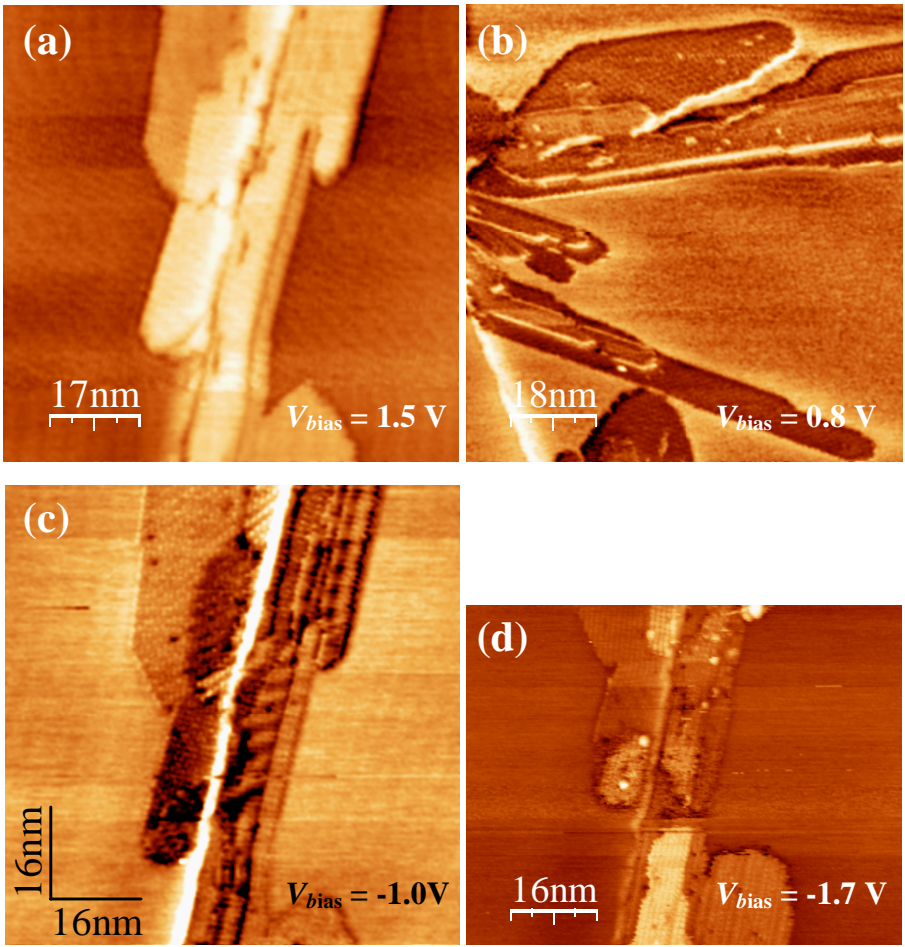

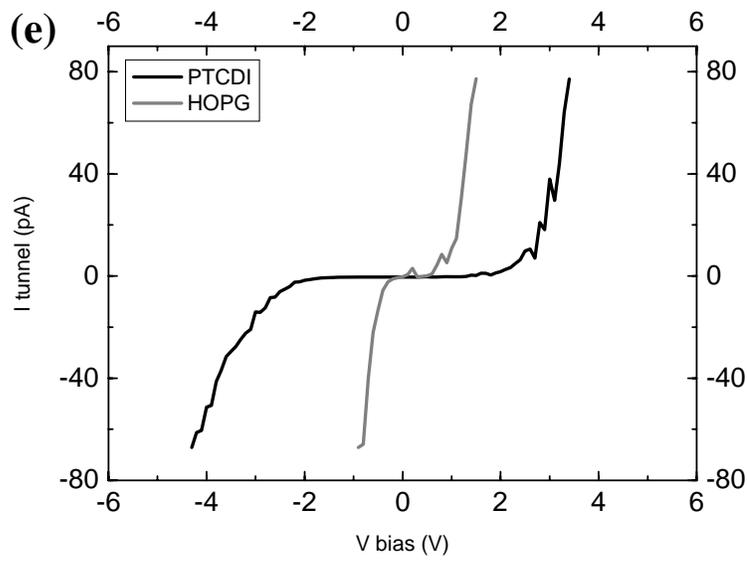

Figure 2

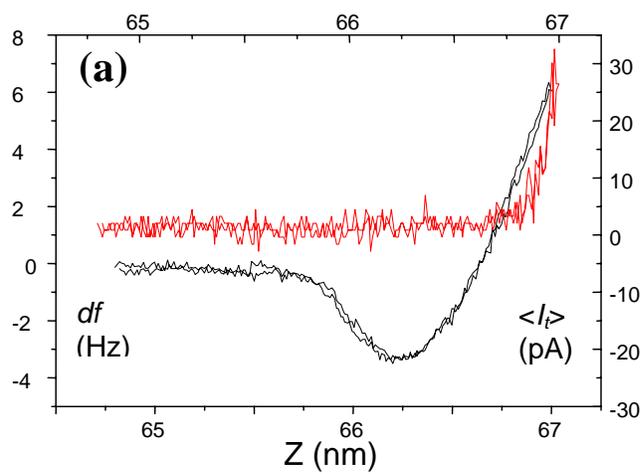
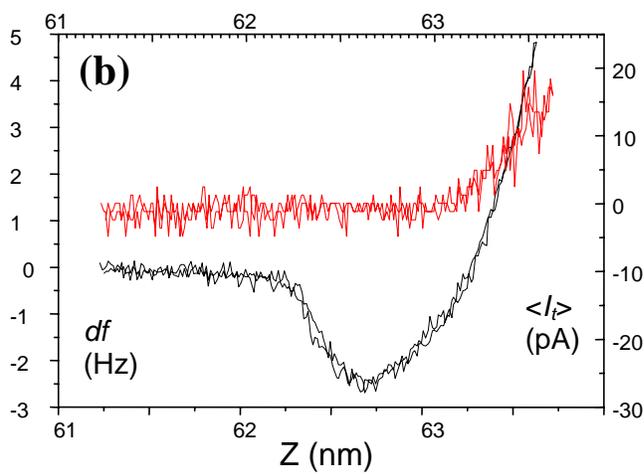

Figure 3